\documentclass[]{article}
\usepackage{amsmath,amssymb,amsfonts}
\usepackage{algorithmic}
\usepackage{graphicx}
\usepackage{textcomp}
\usepackage{xcolor}
\def\BibTeX{{\rm B\kern-.05em{\sc i\kern-.025em b}\kern-.08em
    T\kern-.1667em\lower.7ex\hbox{E}\kern-.125emX}}

\begin{document}

\title{Curving Beam Reflections: Model\\and Experimental Validation\\}

\author{Caroline Jane Spindel and Edward Knightly \\
\textit{Electrical and Computer Engineering} \\
\textit{Rice University}\\
Houston, United States\\
caroline.spindel@rice.edu}

\maketitle

\begin{abstract}
Curving beams are a promising new method for bypassing obstacles in future millimeter-wave to sub-terahertz (sub-THz) networks but lack a general predictive model for their reflections from arbitrary surfaces. We show that, unfortunately, attempting to ``mirror'' the incident beam trajectory across the normal of the reflector, as in ray optics, fails in general. Thus, we introduce the first geometric framework capable of modeling the reflections of arbitrary convex sub-THz curving beams from general reflectors with experimental verification. Rather than ``mirroring'' the trajectory, we decompose the beam into a family of tangents and demonstrate that this process is equivalent to the Legendre transform. This approach allows us to accurately account for reflectors of any shape, size, and position while preserving the underlying physics of wave propagation. Our model is validated through finite element method simulations and over-the-air experiments, demonstrating millimeter-scale accuracy in predicting reflections. Our model provides a foundation for future curving beam communication and sensing systems, enabling the design of reflected curved links and curving radar paths.
\end{abstract}

\section{Introduction}
As the demand for higher data rates in wireless communications and higher resolution in sensing grows, research has increasingly focused on millimeter-wave (mmWave) and terahertz (THz) frequency bands. These bands offer significant advantages in bandwidth and spatial resolution but also introduce fundamental challenges due to their quasi-optical behavior \cite{kurner_thz_2022, shafie_terahertz_2023, chowdhury_6g_2020}. One of the most critical challenges at these frequencies is mitigating blockage events, which can severely degrade system performance \cite{petrov_interference_2017, you_network_2020}.

A promising solution is the use of curving beams, which possess the ability to bend around obstacles \cite{siviloglou_ballistic_2008, singh_wavefront_2024}. A recent study demonstrated that curving beams can recover more power and bit error rate in the presence of blockages than conventional steered beams \cite{guerboukha_curving_2024}. Another study achieved a 400~Gb/s transmission rate using four simultaneous Airy beams \cite{lee_experimental_2025}. Despite growing interest, a critical gap in our understanding of curving beams remains: reflective behavior, which is fundamental to wireless systems such as radar. To our knowledge, previous studies on curving beam reflections have focused on Airy beams interacting with infinite planar dielectric interfaces \cite{yang_reflection_2022, chremmos_reflection_2012, yang_characteristics_2022}. However, real-world scenarios often involve finite reflective surfaces, where curving beam reflections can display more complex and nuanced behavior than these idealized cases suggest. For instance, we will demonstrate that a curving beam impinging upon a small reflector may produce a largely directional, rather than curving, reflection. Yet, aside from directly solving Maxwell's equations, no general model has been proposed to predict the reflection of \textit{arbitrary} curving beams from \textit{arbitrary} surfaces. Moreover, as we will demonstrate, conventional ray optics approaches cannot be applied. Likewise, attempting to ``mirror'' the incident beam trajectory across the normal of the reflector fails in general due to the complex nature of curving beam reflections.

With this work, we aim to bridge the gap between physics and wireless system design. Although curving beam reflections have been largely overlooked in physics, they are crucial for emerging curving communication and sensing systems, where reflections often play a central role in system behavior. Curving beam sensing and communication systems require accurate reflection models, as channel characteristics vary significantly with reflector geometry, as we will show. Our key contributions are as follows:

\textbf{Geometric Reflection Model.} We represent the incident beam trajectory as a family of tangents constrained by the transmitter aperture and reflect those tangents across the local normal of the reflector. We establish that this representation is equivalent to the Legendre transform. The key insight is that the set of reflected tangents encodes the beam's post-reflection path, enabling the construction of its reflected trajectory. Our method does not ``mirror'' the entire incident beam, as such an approach would fail for both finite and nonplanar reflectors. Likewise, we do not perform complex electric field analysis via Maxwell's equations. In contrast, our approach allows us to accurately account for any convex beam and reflectors of any shape, size, and position while preserving the underlying physics of wave propagation. 

\textbf{Simulation Validation.} We validate our model using finite element method (FEM) simulations across multiple scenarios, specifically large planar, small planar, and convex reflectors. We demonstrate that providing a single function representing the trajectory of the beam’s peak intensity is sufficient to accurately predict the reflected peak intensity, even when the reflector produces a directional beam. Additionally, using two functions to describe the beam’s left and right edges allows us to reconstruct the full trajectory of the reflected lobe - an approach that is particularly effective for nonplanar reflectors. 

\textbf{Experimental Validation.} We further validate our model via over-the-air experiments using a THz time-domain spectroscopy (TDS) system. We first generate a parabolic curving beam using a 3D-printed phase plate and characterize the beam's curvature prior to reflection. We then introduce a large planar metal reflector and measure the resulting reflection. We observe strong agreement between the experimental results and our model’s predictions. When predicting the peak of the reflected beam, our model achieves errors as low as 7 mm. 

The remainder of the paper is organized as follows: Section~\ref{background} provides background on self-curving beams. We introduce our model in Section~\ref{model} and perform the simulation evaluation in Section~\ref{simulations}. Section~\ref{experiments} provides a description of our experimental setup and results. We discuss related work in Section~\ref{relatedwork} and conclude the paper in Section~\ref{conclusion}.

\section{BACKGROUND} \label{background}
\begin{figure}[tb]
    \centering
    \includegraphics[width=\columnwidth]{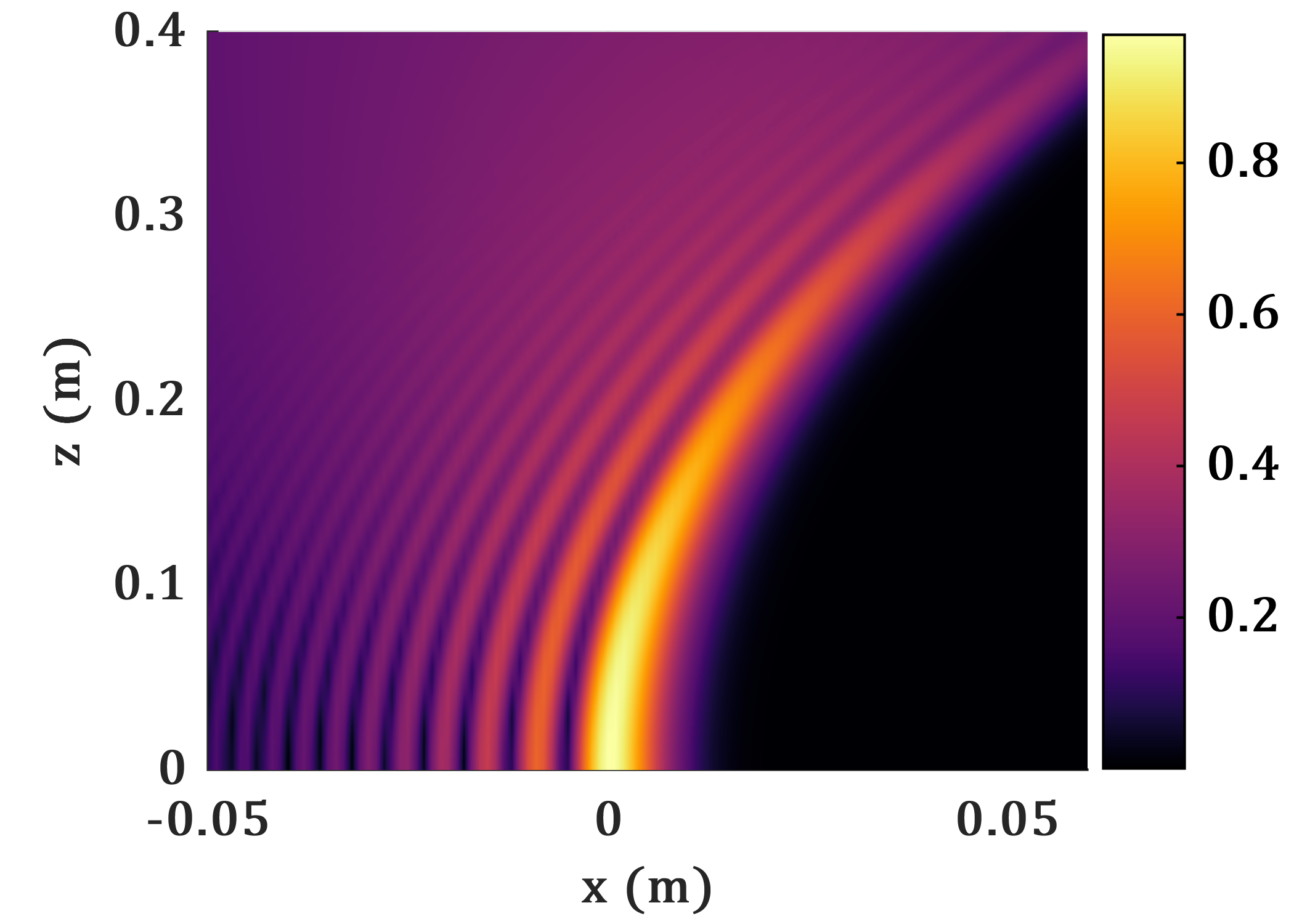}
    \caption{Electric field distribution of a truncated Airy beam at 150 GHz.}
    \label{fig:airybeam}
\end{figure}

In this section, we provide a brief overview of curving beams. We use the Airy beam as a representative example, given its well-defined and extensively studied electric field~\cite{berry_nonspreading_1979, siviloglou_observation_2007, siviloglou_accelerating_2007, efremidis_airy_2019, siviloglou_ballistic_2008, sztul_poynting_2008, rogel-salazar_full_2014}. However, the model presented in this work is not limited to Airy beams; it is broadly applicable to any convex curving beam, including those discussed in~\cite{froehly_arbitrary_2011,greenfield_accelerating_2011, guerboukha_curving_2024}. The electric field of a finite energy Airy beam can be expressed as:
\begin{equation}
\label{eq:airy}
\begin{split}
    E(\xi,s) =\; & \mathrm{Ai} \bigg( s - \bigg( \frac{\xi}{2}^2 \bigg) +ia\xi \bigg) \cdot \\
    & \exp \bigg(as - \frac{a\xi^2}{2} - i\frac{\xi^3}{12}+ i\frac{a^2\xi}{2} + i\frac{s\xi}{2} \bigg)
\end{split}
\end{equation}
where $x$ is the transverse axis, $z$ is the propagation axis, $\text{Ai}$ is the Airy function, $\xi=z/kx_0^2$ is the propagation distance normalized, $x_0$ is an arbitrary transverse scale related to the full width at half maximum of the main lobe, $s=x/x_0$ is the scaled transverse coordinate, $k=2\pi f/c$ is the wavenumber, and $a$ is a factor accounting for the truncation of the beam as it propagates~\cite{siviloglou_accelerating_2007, siviloglou_observation_2007}. From Equation~\eqref{eq:airy} we can extract the term $z^2/4k^2x_0^3$ from the Airy function. This term governs the beam's parabolic trajectory and reveals that its self-bending behavior is primarily influenced by the frequency (through the wavenumber $k$) and $x_0$. We plot a 150~GHz Airy beam in Fig.~\ref{fig:airybeam}.
For any convex curving beams, not just Airy beams, the size of the emitting aperture defines both the beam's trajectory and the distance over which it can propagate without substantial truncation~\cite{guerboukha_curving_2024}.

\section{Modeling Curving Beam Reflections} \label{model}
\subsection{Model Overview}
The key contribution of this work is a geometric model for predicting how any convex curving beam reflects off perfectly conductive reflectors with no wavelength-scale features. This presents a fundamental challenge as conventional ray tracing methods assume straight-line propagation and do not account for beam curvature. A natural alternative is to ``mirror'' the trajectory of the beam across the surface normal; however, this method is \textit{only} effective for very large planar reflectors. A curving beam impinging upon a finite reflector may exhibit reduced curvature after reflection, which ``mirroring'' the trajectory cannot account for. Moreover, it is entirely unsuitable for predicting the reflections from nonplanar surfaces. We exclude rough reflectors from our model for two reasons: first, the physics of sub-THz wave interactions with rough surfaces remains an active research area; and second, to our knowledge, the reflection behavior of engineered sub-THz wavefronts - regardless of beam shape - has not been explored. While this lies beyond the scope of the present work, it represents an important direction for future investigation.

Instead of attempting to ``mirror'' the trajectory of the incident beam in its entirety, we decompose the trajectory into a family of tangent lines. This decomposition is equivalent to the Legendre transform. The Legendre transform provides an alternative representation of a convex function in terms of its derivative. The Legendre transform $F^*$ of the function $F$ can be written as
\begin{equation}
\label{eq:lt}
F^*(m) \equiv mx(m)-F(x(m))
\end{equation}
where $x(m)$ is obtained by inverting the slope function $m(x)=dF(x)/dx$ \cite{zia_making_2009}. The function $F$ becomes an envelope for a family of tangents. 

By decomposing the incident beam trajectory into its corresponding family of tangents, we can apply the laws of reflection to each tangent \textit{individually} rather than to their entire envelope. This approach allows us to account for finite reflectors by considering only tangents that intersect the reflector. Furthermore, we can account for nonplanar reflectors by reflecting each tangent according to the local surface normal at the point of intersection between the tangent and the reflector. After reflecting the tangents, we can then construct the reflected beam using the invertibility of the Legendre transform. If no envelope exists or if the envelope approximates a linear function, this indicates a directional reflection rather than a curving reflection.

In general, the tangent decomposition of an arbitrary convex curve is most conveniently performed through numerical methods, particularly when the solution must be bounded by a finite aperture. Thus, the tangent decomposition presented in the following section can be regarded as a numerical implementation of the Legendre transform.

\begin{figure}[tb]
    \centering
    \includegraphics[width=\columnwidth]{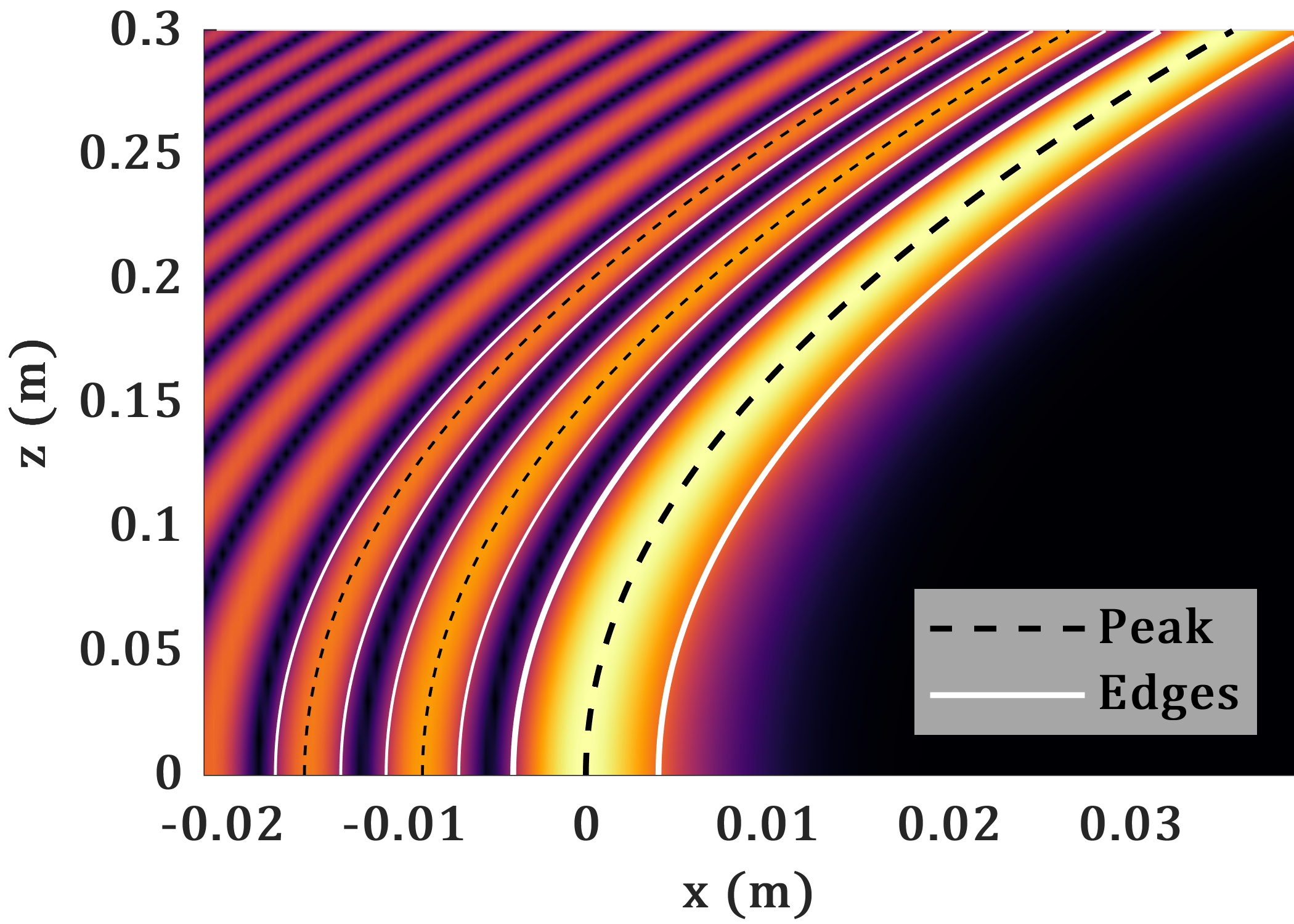}
    \caption{Characterizing the trajectory of an Airy beam.}
    \label{fig:trajchara}
\end{figure}

\begin{figure}[tb]
    \centering
    \includegraphics[trim=200 20 200 20, clip, width=\columnwidth, height=2.5in, keepaspectratio]{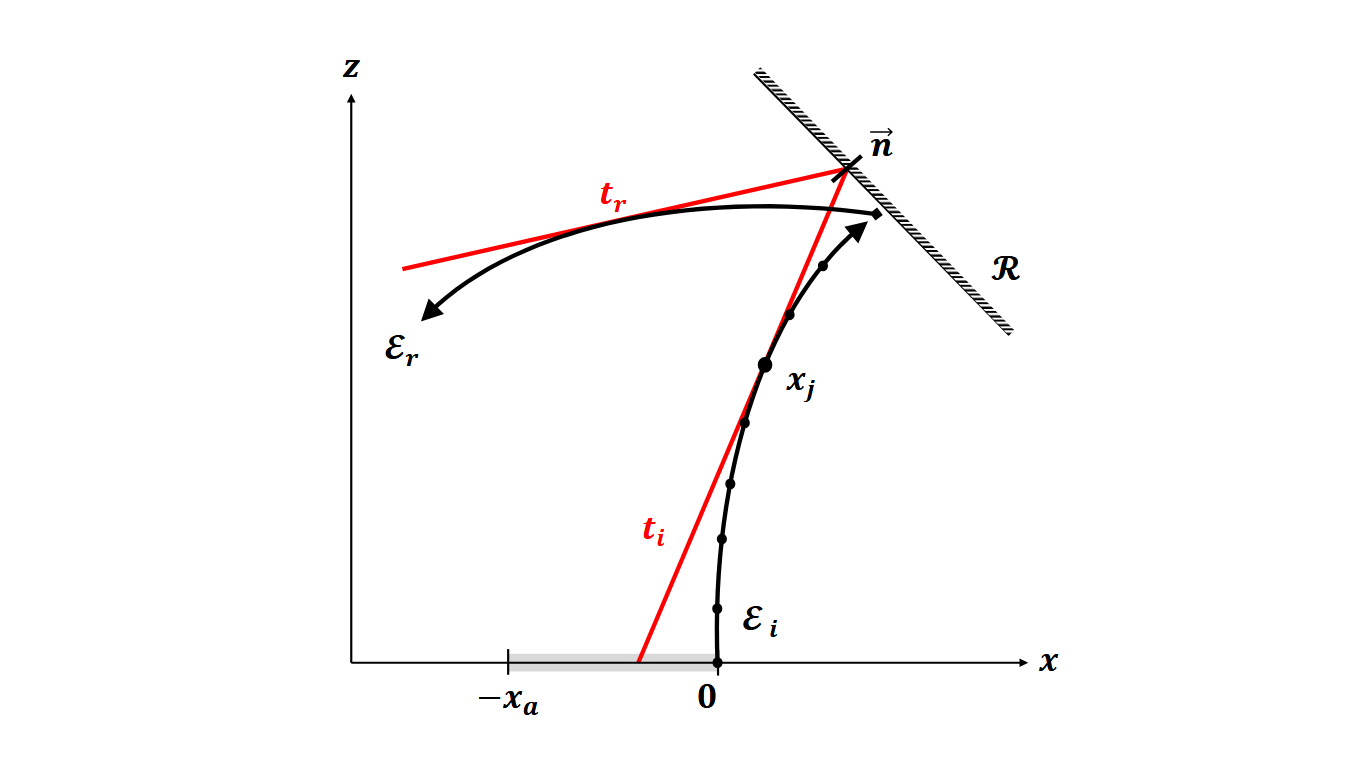}
    \caption{An illustration of the model's inputs and outputs.}
    \label{fig:fmodel}
\end{figure}

\subsection{Model Inputs} \label{sec:fmodel_inputs}
The model considers convex curving beams observed on the $xz$-plane, as the Legendre transform is most effective when applied to convex functions. While this framework can, in principle, be extended to nonconvex trajectories, doing so requires more careful handling of the Legendre transform~\cite{zia_making_2009}. We define the $x$-axis as the transverse direction and the $z$-axis as the propagation direction. We assume that the beam is invariant along the $y$-axis. The transmitter aperture is planar and spans the range $x\in[-x_a, 0]$ at $z=0$. 

To characterize the incident beam, we define its trajectory as a function $\mathcal{E}_i$, which serves as the envelope to the family of tangents in the Legendre transform. In many cases, $\mathcal{E}_i$ corresponds to the trajectory along which the peak intensity of the beam is designed to propagate. Given the peak trajectory as input, the model predicts the peak trajectory of the reflected beam. Throughout this paper, we refer to this as a ``single-trajectory'' input.

For Airy beams, the incident peak trajectory is given by:
\begin{equation}
\label{trajz}
    \mathcal{E}_i(z)=\frac{z^2}{4k^2x_0^3}
\end{equation}
which we derive from Equation~\eqref{eq:airy}. 

In certain cases, it is important to account for how reflection affects the beamwidth. For example, a curving beam incident upon a convex reflector may become more directional and broaden in width after reflection. To capture this behavior, we extend the beam characterization by defining its left and right edges, defined as $\mathcal{E}_i(x \pm \text{beamwidth})$. For an Airy beam, this becomes $\mathcal{E}_i(x \pm x_0)$. The model processes each edge independently: the trajectory corresponding to the left edge of the incident beam produces the left edge of the reflected beam, and likewise for the right edge. Throughout the paper, we refer to this as ``multiple-trajectory'' input.

Some curving beams exhibit multiple lobes, and analyzing the reflection of these lobes may provide a more comprehensive prediction. Consider the case of a multi-lobed curving beam impinging upon a convex reflector: the lobes may spatially diverge after reflection. Similar to the beamwidth-based characterization, the model can independently analyze each lobe. Fig.~\ref{fig:trajchara} illustrates this approach for a 150~GHz Airy beam. Dashed lines represent beam characterization based on the trajectory of the peak. Solid lines represent characterization using the left and right beam edges.

Because the model processes each input trajectory independently, the number of outputs directly corresponds to the number of inputs. Modifying one input does not affect the predictions for others - each trajectory is evaluated in isolation. For example, if three distinct input trajectories are provided, the model will generate three corresponding reflected trajectories. This property ensures that the model maintains an independent evaluation of each input trajectory, making it adaptable to complex multi-lobed beam structures.

Lastly, the model requires the reflector profile $\mathcal{R}$ as an input. The function $\mathcal{R}$ defines the two-dimensional shape of the reflecting surface and may be any arbitrary real-valued function, finite or infinite in extent. As with any physics-based model, the accuracy of this approach depends on how precisely $\mathcal{R}$ captures the true reflector geometry - an inherent and expected tradeoff in modeling physical behaviors. Fig.~\ref{fig:fmodel} illustrates the overall setup. A convex curving beam trajectory $\mathcal{E}_i$ impinges on a reflector $\mathcal{R}$, resulting in a reflected envelope $\mathcal{E}_r$. The transmitter aperture is positioned at $z=0$ and spans the range $x\in[-x_a,0]$.

\subsection{Decomposing the Trajectory into Tangents}
We first decompose the trajectory into a family of tangents. For ease of exposition, we illustrate this process using the single-trajectory characterization of the incident beam and consider it as a function of $x$.

We perform the Legendre transform numerically, as illustrated in Fig~\ref{fig:fmodel}. We sample $j$ equally spaced points along the trajectory $\mathcal{E}_i(x)$, ranging from the point of intersection with the $z$-axis to the farthest propagation point determined by the aperture size $x_a$, denoted as $(x_{max}, \mathcal{E}_i(x_{max}))$~\cite{guerboukha_curving_2024}. At each sampled point $x_j$, we compute the slope of the corresponding tangent line as
\begin{equation}
\label{eq:slope}
    m_{i,j} = \frac{d \mathcal{E}_i(x)}{dx} \big|_{x=x_j}
\end{equation}
The corresponding $z$-intercept is given by
\begin{equation}
    \psi_j=\mathcal{E}_i(x_j)-m_{i,j}x_j
\end{equation}
Thus, the equation of the $j^{th}$ tangent line can be expressed as
\begin{equation}
    t_{i,j}=xm_{i,j} + \psi_{i_j}
\end{equation}
The vector describing the $x$ and $z$ direction of the tangent as a ray can be expressed as
\begin{equation}
\label{eq:tdir}
    t_{i,j}= 
    \begin{bmatrix}
    1 \\ m_{i,j}
    \end{bmatrix}
\end{equation}
This vector form simplifies subsequent reflection calculations. Increasing the number of tangents $j$ increases the resolution of the reflection prediction with diminishing returns. Therefore, the number of tangents should be scaled proportionally to the aperture size.

\subsection{Reflecting the Tangents}
We identify the point where each tangent intersects with the reflector $\mathcal{R}$. At this intersection point, we calculate the unit normal vector $\vec{n}$ to the reflector surface. In the case of a planar reflector, the normal vector remains constant across all intersection points. In the case of a nonplanar reflector, the normal will vary according to the surface geometry, which we must account for in the reflection calculation. To account for finite reflectors, we exclude any incident tangents that do not intersect with the reflector.

We compute the reflection of each tangent using the law of reflection, viz., the angle of reflection of a ray, measured relative to the surface normal, is equal to the angle of incidence. For the $j^{th}$ incident tangent, given the direction vector $\vec{t}_{i,j}$ and the unit surface normal at the point of reflection $\vec{n}_j$, the direction vector of the reflected tangent $\vec{t}_{r,j}$ is computed as:
\begin{equation}
\label{eq:lor}
\vec{t}_{r,j}=\vec{t}_{i,j}-2(\vec{t}_{i,j} \cdot \vec{n}_j)\vec{n}_j
\end{equation}
which ensures that the reflected tangent maintains the correct angle with respect to the surface, regardless of the reflector's shape. The slope of the reflected ray will be
\begin{equation}
    m_{r,j}=\frac{\vec{t}_{r,j,z}}{\vec{t}_{r,j,x}}
\end{equation}
where $\vec{t}_{r,j,x}$ and $\vec{t}_{r,j,z}$ are the $x$- and $z$-components of the reflected tangent, respectively. The equation of the reflected tangent line can then be expressed using the intersection point of the incident ray with the reflector. 

\subsection{Reconstructing the Reflection}
The reflected tangent rays form a family of tangents, which define the envelope of the reflected beam. There are several established techniques for deriving the equation of the envelope from a given family of tangents. In this work, we use a straightforward numerical method: we identify discrete points along the apparent edge defined by the reflected tangents and fit a curve that best approximates the resulting trajectory.

\begin{figure}[tb]
  \centering
    \includegraphics[width=\linewidth]{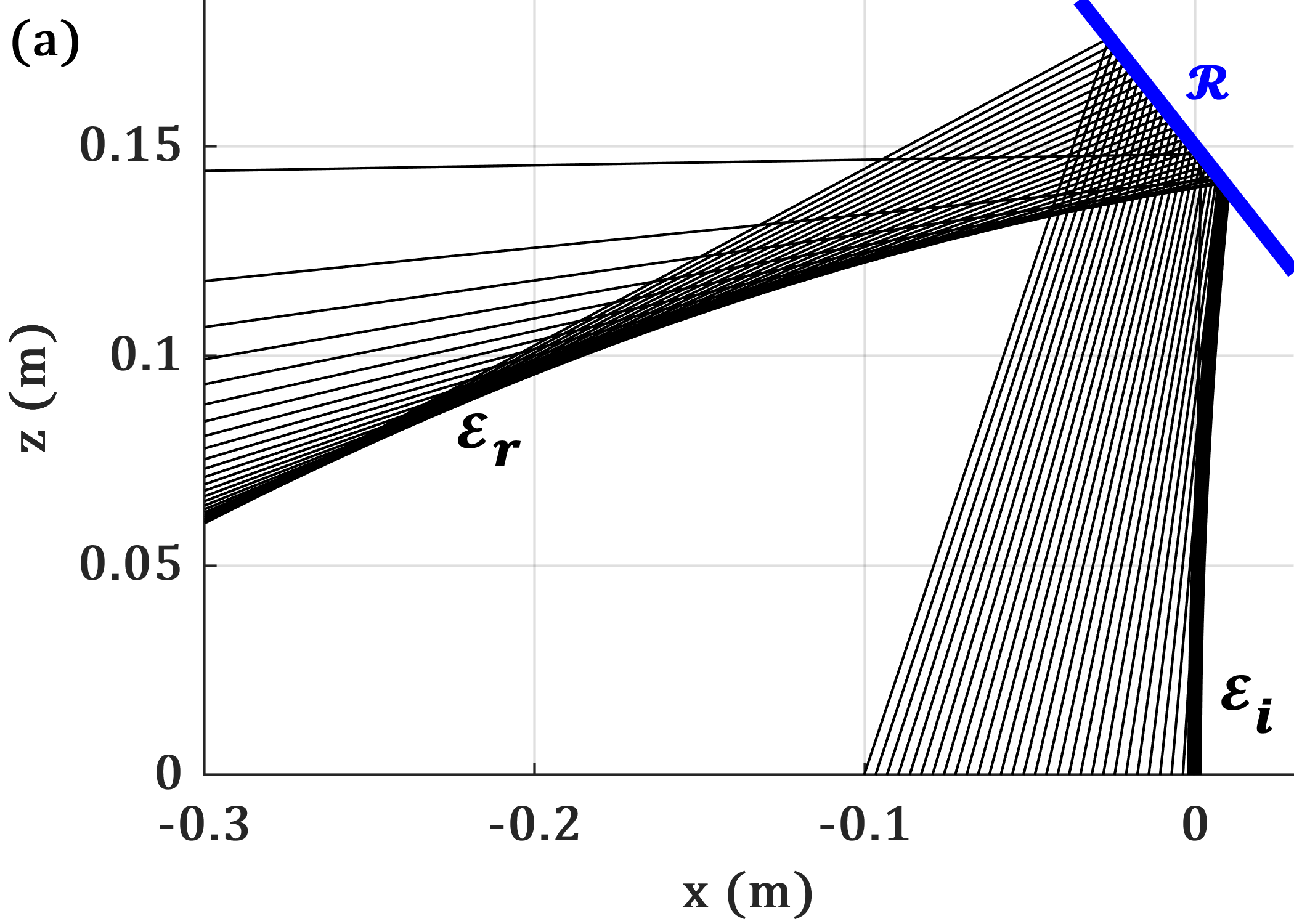}
    \includegraphics[width=\linewidth]{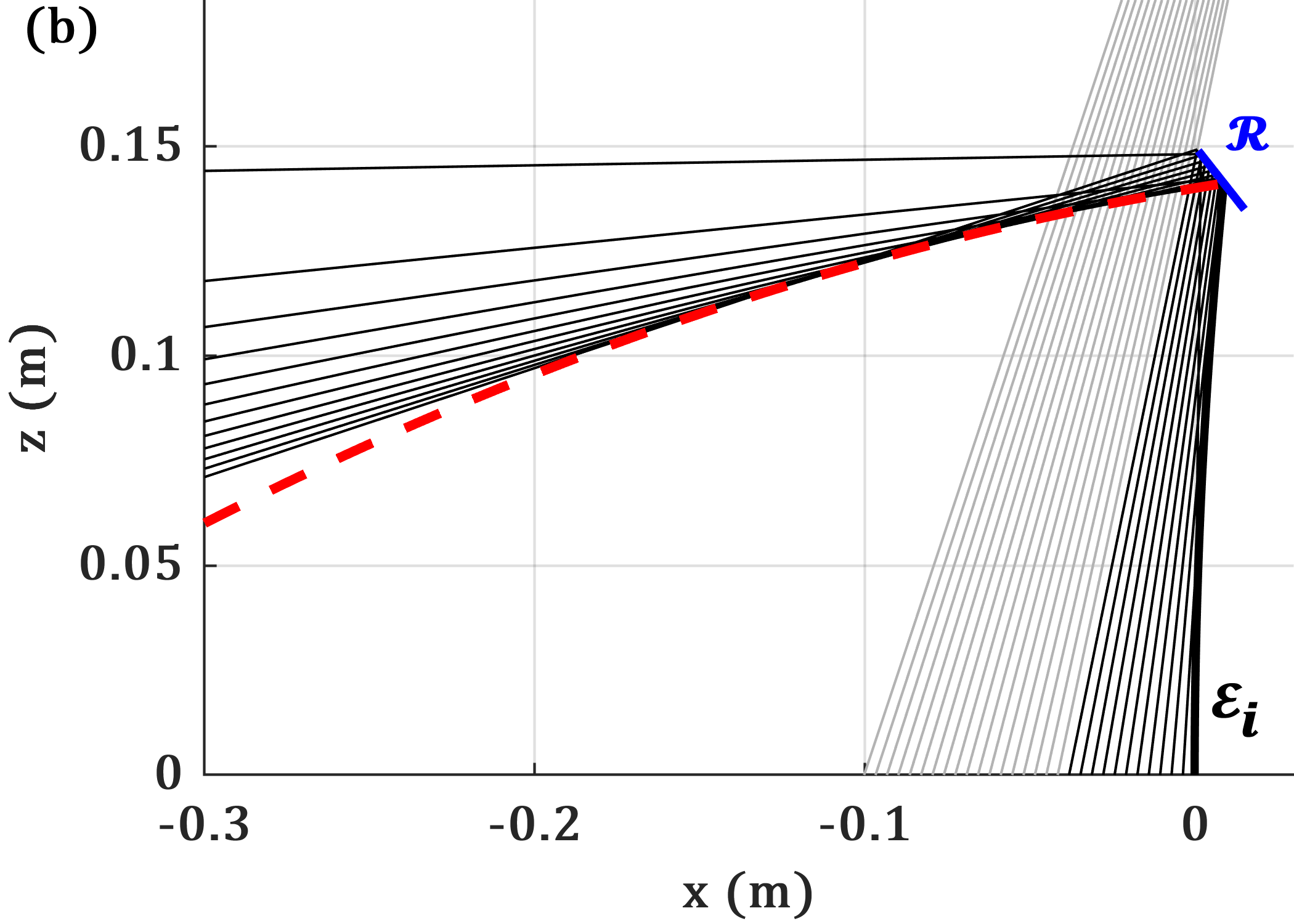}
  \caption{Predicting the reflection off (a) a large planar reflector and (b) a small planar reflector.}
  \label{fig:bigsmall}
\end{figure}

In Fig.~\ref{fig:bigsmall}(a), we demonstrate the model's prediction of a parabolic incident beam with the trajectory envelope $\mathcal E_i(x)=1.5\sqrt{x}$ reflecting off an infinite planar reflector described by $\mathcal R(x)=-x+0.150$. We set the aperture to $x_a=0.1$. Using these inputs, we calculate $j=30$ tangents and reflect them across the surface of the reflector. Because the reflector is infinite and planar, the reflected tangents form a cohesive curving envelope. The model, therefore, predicts that the reflected beam follows a curving trajectory described by $\hat{\mathcal{E}_r} (x)=-0.4262x^2+0.1374x+0.1406$.

In contrast, when no clear envelope emerges from the reflected tangents or the resulting envelope approximates a linear function, the model predicts that the reflection is directional rather than curving. We illustrate this scenario in Fig.~\ref{fig:bigsmall}(b), where the same incident beam reflects off the same reflector, but the reflector is now finite and restricted to the interval $x\in[1, 15]$ mm. Despite using the same number of tangents, the reflected rays fail to form a coherent curving envelope. As a result, the model predicts that the reflected beam will be a directional beam diverging in width. In contrast, ``mirroring'' $\mathcal{E}_i$ across the normal of this reflector would incorrectly suggest a reflected \textit{curving} beam, as depicted by the red dashed line in Fig.~\ref{fig:bigsmall}(b).

Some curving beams exhibit prominent sidelobes (see Fig.~\ref{fig:airybeam}), which may also be important to capture when modeling reflections. Understanding how the reflector affects the beam's width can also be of interest. To explore these effects, we again consider the same incident beam and infinite planar reflector as in previous examples. 

\begin{figure}[tb]
    \centering
    \includegraphics[width=\columnwidth]{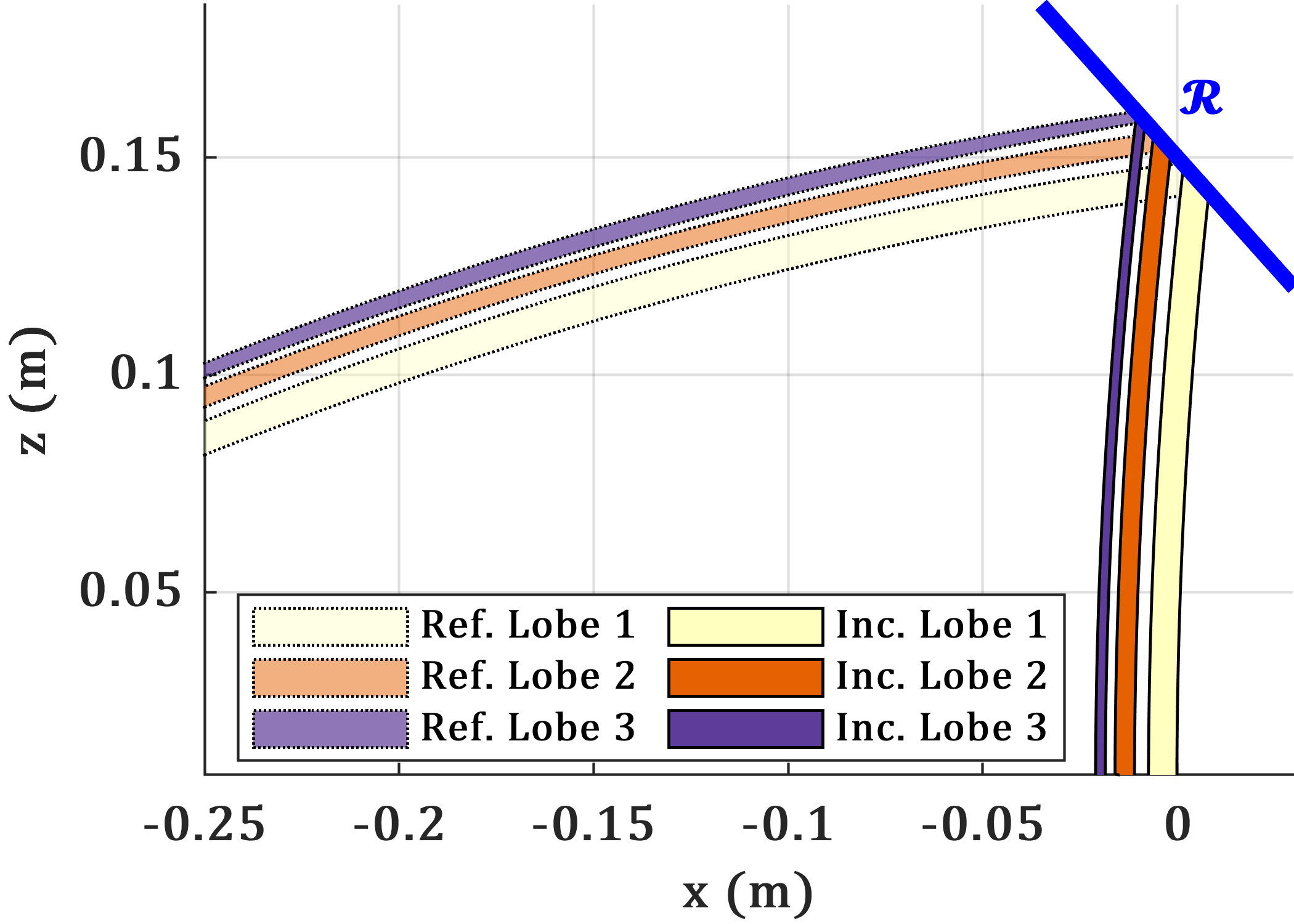}
    \caption{Prediction accounting for the beamwidth and three lobes (tangents not shown).}
    \label{fig:lobes}
\end{figure}

When using multiple inputs to define the right and left edges of the incident beam, the model applies the same tangent decomposition and reflection process independently to each input. In Fig.~\ref{fig:lobes}, we demonstrate this with the same incident beam as in Fig.~\ref{fig:bigsmall}. However, rather than using a single trajectory describing the peak intensity, we now provide the left and right edges of the first three lobes as inputs. We perform trajectory decomposition and reflection reconstruction for each input, yielding the prediction shown in Fig.~\ref{fig:lobes} (the individual tangents have been excluded from the figure for clarity). The model accurately predicts that each lobe will curve post-reflection while maintaining its width, consistent with the expected behavior when reflecting off an infinite planar reflector.

\subsection{Computational Complexity}
Predicting reflected beam behavior using Maxwell's equations with methods like FEM requires discretizing the simulation space and solving at each point, leading to trade-offs between accuracy and speed. These simulations are computationally intensive and often lack interpretability. In contrast, our model considers only the relevant incident beam trajectory, computing a single derivative that is reused to determine tangent slopes. Our (unoptimized) MATLAB implementation of the model runs approximately 100 times faster than a comparable FEM simulation. Beyond speed, our model also provides insight into reflection behavior - an advantage often missing in full-wave simulations.

\section{MODEL EVALUATION WITH FEM SIMULATIONS}
\label{simulations}
In this section, we evaluate the proposed model with FEM simulations, which numerically solve Maxwell’s equations under defined initial and boundary conditions. We assess our model’s ability to predict the reflected beam given the incident beam and reflector geometry. All FEM simulations are conducted using the COMSOL Multiphysics platform, and the model calculations are implemented in MATLAB. 

\subsection{Scenario and Performance Metrics}
We use the same incident beam throughout all evaluation scenarios: a 150 GHz Airy beam with $x_0=3.6$ mm and a truncation factor of $a = 0.1$. The aperture size $x_a$ is 200 mm. All reflectors are simulated as perfect conductors.

We examine four cases: (1) a single-trajectory input and a large planar reflector; (2) multiple-trajectory input and a large planar reflector where beamwidth and multiple lobes are taken into account; (3) a single-trajectory input and a small planar reflector; and (4) multiple-trajectory input and a convex reflector where multiple lobes are taken into account. These scenarios represent common reflection behaviors. For instance, irregular reflectors often produce behavior similar to the convex reflector.

To evaluate the accuracy of the model in scenarios where the reflection produces a measurable envelope, we compute the root mean square error (RMSE) between the actual reflected envelope $\mathcal{E}_r$ and the predicted envelope $\hat{\mathcal{E}}_r$, calculated as
\begin{equation}
\text{RMSE}=\sqrt{\frac{1}{n}\sum_{i=1}^{n}\left|\mathcal{E}_{r,i} - \hat{\mathcal{E}}_{r,i}\right|^2}
\end{equation}
We calculate the maximum distance between actual and predicted values as
\begin{equation}
    \max{|\mathcal{E}_r(x)-\hat{\mathcal{E}}_r(x)|}
\end{equation}

\begin{figure*}[tb]
  \centering
  \begin{tabular}{cc}
    \includegraphics[width=0.45\linewidth]{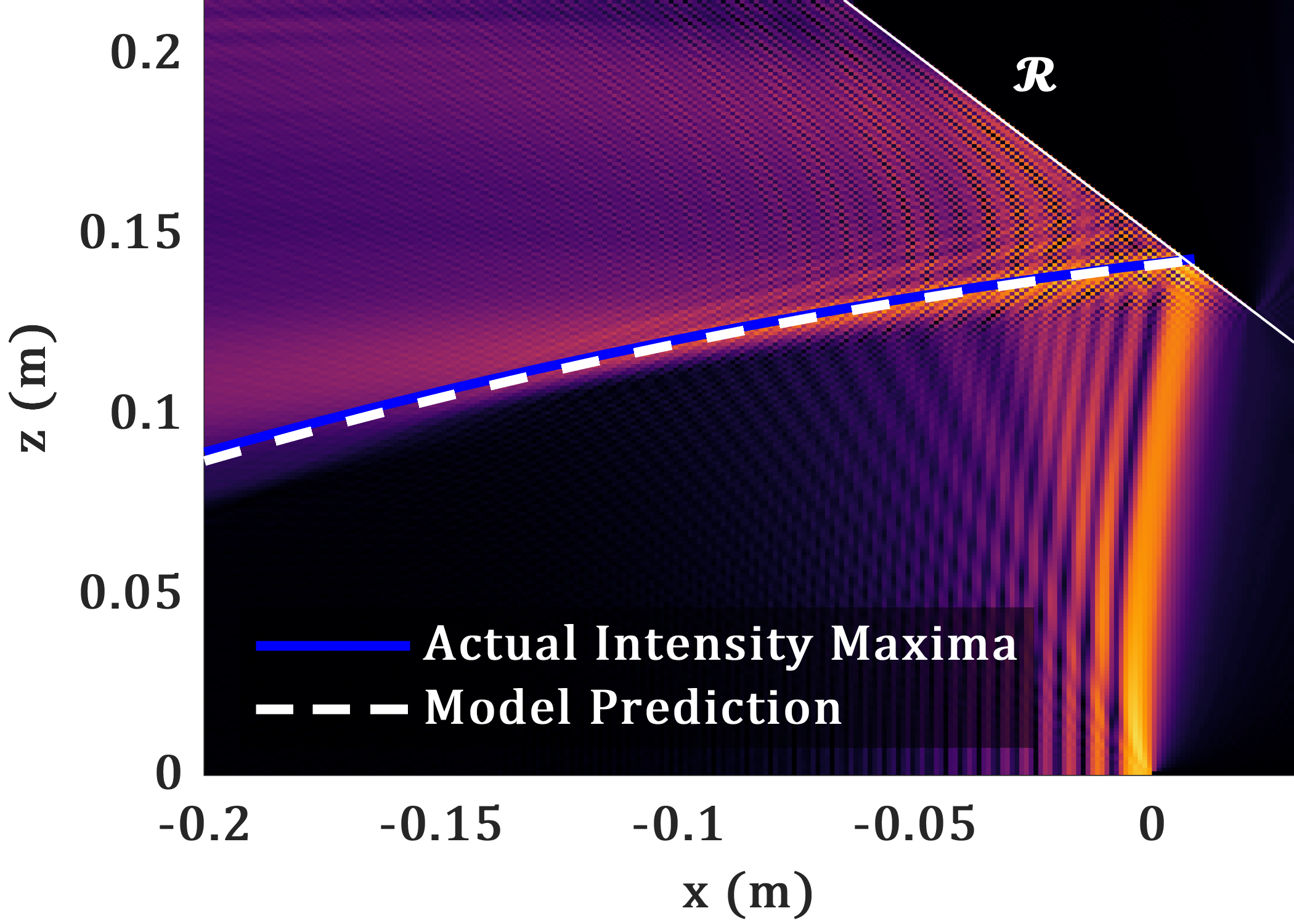} &
    \includegraphics[width=0.45\linewidth]{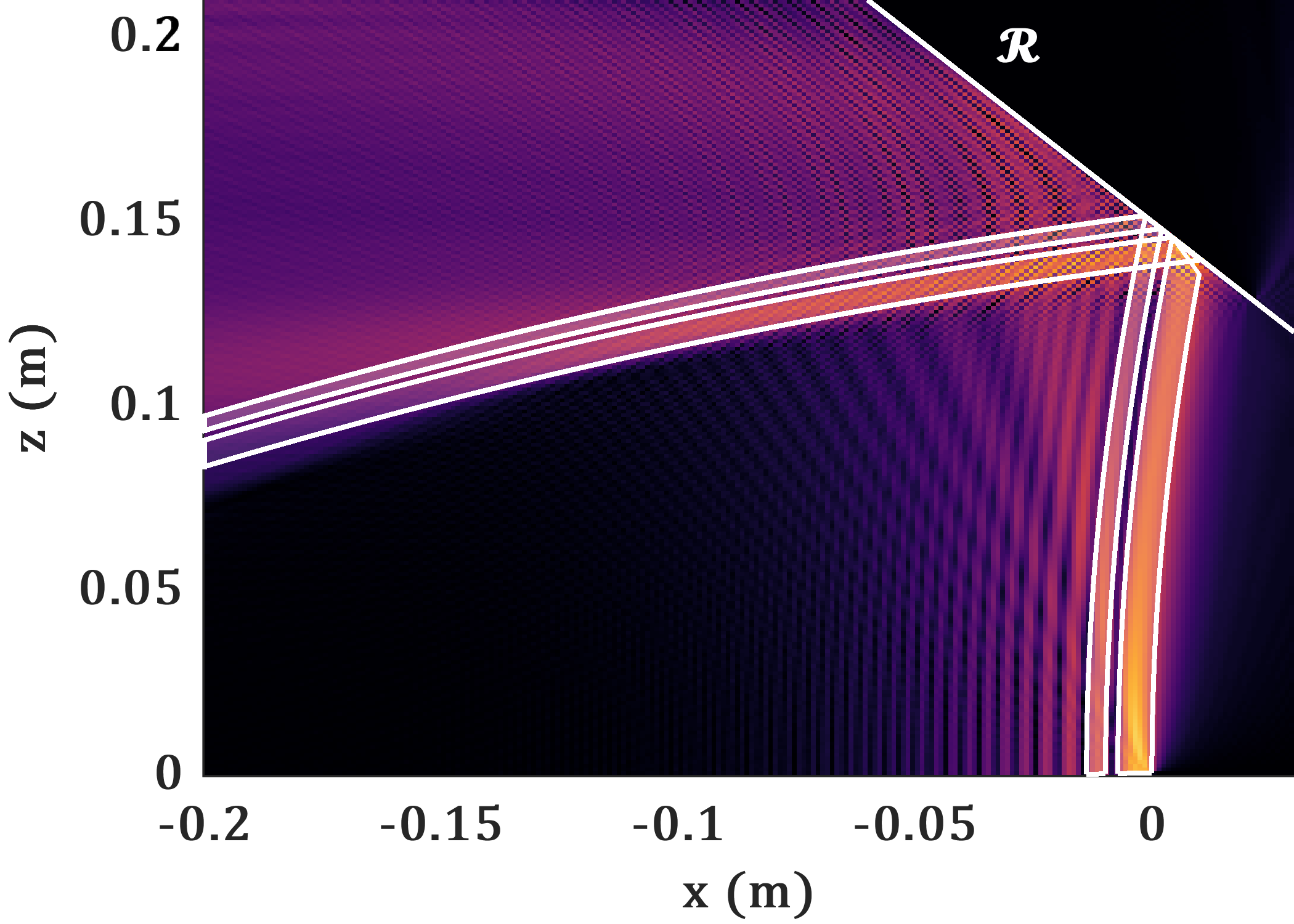} \\
    {\footnotesize (a) Single-trajectory input with a large planar reflector.} & {\footnotesize (b) Multiple-trajectory input with a large planar reflector.}  \\ \\
    \includegraphics[width=0.45\linewidth]{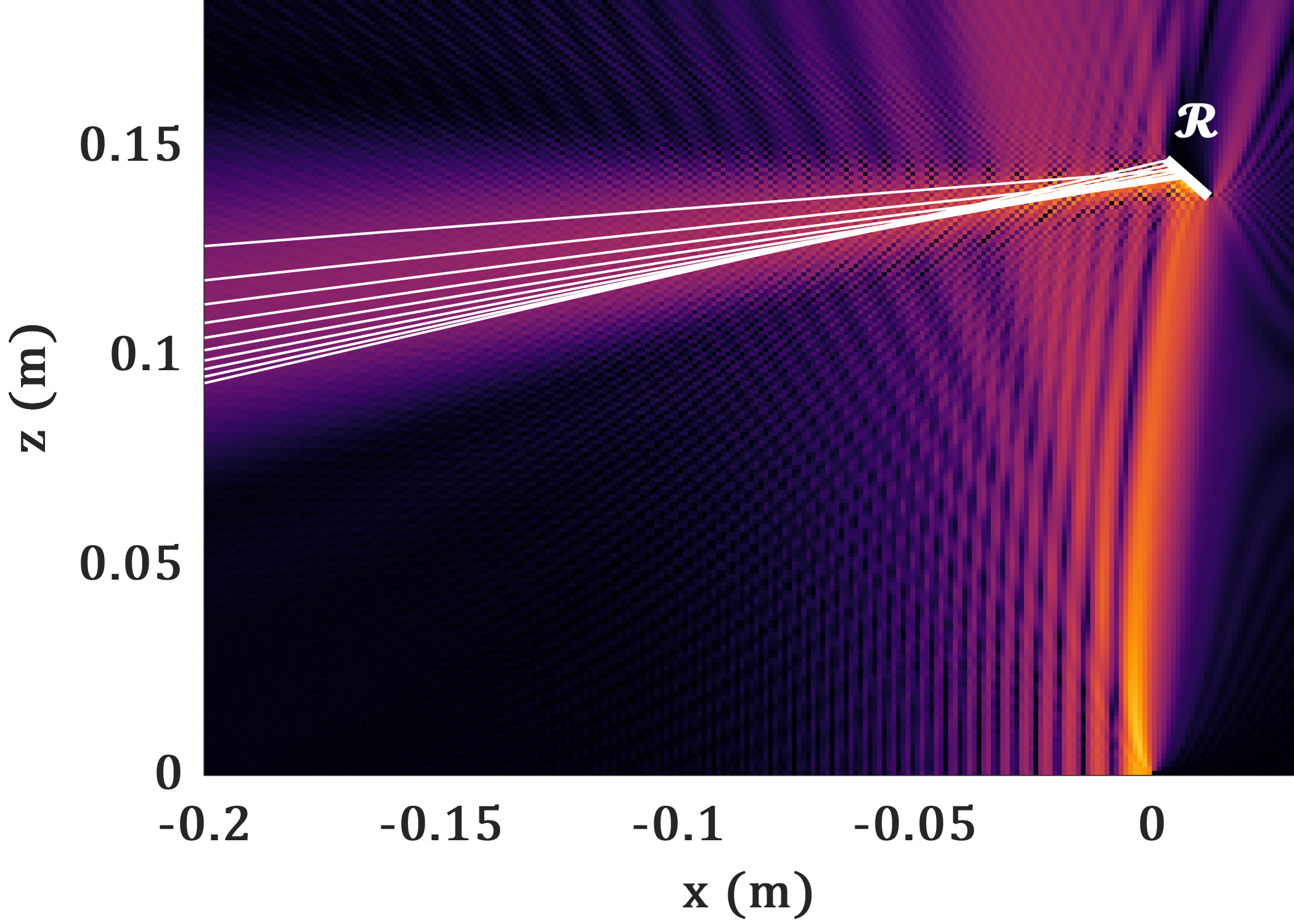} &
    \includegraphics[width=0.45\linewidth]{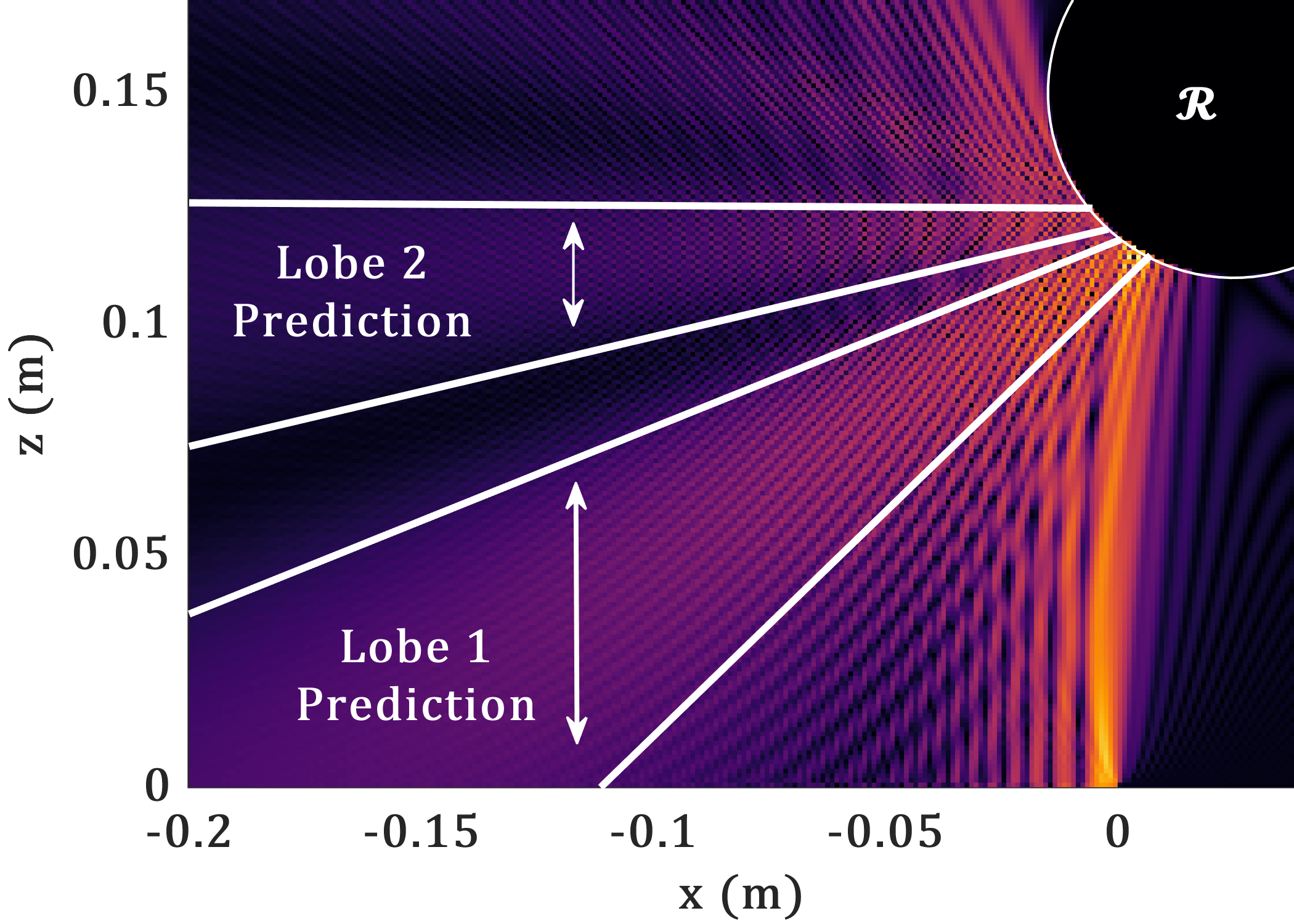} \\
    {\footnotesize (c) Single-trajectory input with a small planar reflector.} & {\footnotesize (d) Multiple-trajectory input with a convex reflector.} \\
  \end{tabular}
  \caption{FEM simulation validation of four reflection scenarios.}
  \label{fig:fem}
\end{figure*}

\subsection{Single-Trajectory Input with Large Planar Reflector}
\label{sec:fmodel_s1}
In this scenario, we evaluate the model's ability to predict a curving reflection given a single trajectory as input. The reflector function $\mathcal{R}(x)=-x+0.150$ is bounded between $x\in[-20, 65]$ mm, which is a sufficient size to produce a curving reflection. To visualize the simulation output, we generate a heatmap of the electric field intensity (Fig.~\ref{fig:fem}(a)) and find that the reflection forms a curving envelope due to the large planar reflector. We extract the peak electric field intensity, which follows the function $\mathcal{E}_r(x)=-0.5192x^2 + 0.1577x + 0.1418$. This trajectory is shown as a solid blue line in the figure.

We provide the model with three inputs: the aperture, the reflector function $\mathcal{R}(x)$, and the trajectory of the first lobe along its intensity maxima. Given these inputs, the model predicts that the reflection maintains its curvature. The predicted reflected envelope follows: $\hat{\mathcal{E}_r}(x)= -0.5406x^2 + 0.1635x + 0.1414$, which is represented as a white dashed line in Fig.~\ref{fig:fem}(a). For clarity, the individual tangents are omitted. 

To assess prediction accuracy, we compute the RMSE between the predicted and simulated reflection trajectories, obtaining a value of 1.4 mm, which is smaller than the wavelength. The maximum distance between $\hat{\mathcal{E}_r}(x)$ and $\mathcal{E}_r(x)$ is 2.4 mm. These results confirm that the model provides accurate predictions for curving reflections off large planar surfaces with a single-trajectory input.

\subsection{Multiple-Trajectory Input with Large Reflector}
\label{sec:fmodel_s1_m}
We expand upon the setup described in Section~\ref{sec:fmodel_s1} and use the same simulation to evaluate the model's ability to predict the behavior of multiple lobes and their beamwidths after reflection. In the previous scenario, the model predicted only the intensity maxima of the first lobe reflected. Here, we examine how multiple lobes interact with a planar reflector and how their spatial characteristics evolve post-reflection. 
 
In addition to the aperture and the reflector function, we now input the left and right edges of the first two lobes for a total of four functions. These edges define both the beamwidth and the spatial separation of the two lobes, enabling the model to track their position and beamwidth after reflection. Consequently, the model produces four output functions, corresponding to the right and left edges of the reflected lobes, as illustrated in Fig.~\ref{fig:fem}(b). For clarity, the individual tangents are omitted. The model accurately predicts that the reflected lobes maintain their curvature, spatial separation, and beamwidth post-reflection. As shown in the figure, the predicted lobes closely match the expected shape and spatial properties. 

\subsection{Single-Trajectory Input with Small Planar Reflector}
In this scenario, we assess the model's ability to predict a directional reflection given a single-function input. We bound $\mathcal{R}(x)=-x+0.150$ between $x\in[3, 12]$ mm, ensuring a directional reflection. We visualize the simulation output as a heatmap in Fig.~\ref{fig:fem}(c). As expected, the reflection is directional with no curving envelope.

We provide the model with three inputs: the aperture size, $\mathcal{R}(x)$, and $\mathcal{E}_i(x)$. Unlike the previous scenarios, the reflected tangents do not form a coherent envelope, indicating a directional reflection rather than a curving one. As shown in Fig.~\ref{fig:fem}(c) with the reflected tangents as white lines, the model correctly captures this behavior, and the upper and lower tangents align with the corresponding edges of the reflected beam, accurately predicting its width and directionality.

This scenario further highlights the limitations of the ``mirroring'' approach for predicting the reflected trajectory. If we simply ``mirrored'' the incident trajectory $\mathcal{E}_i(x)$ across the normal of the reflector profile $\mathcal{R}(x)$, we would erroneously predict that the reflection is curving rather than directional.

\subsection{Multiple-Trajectory Input with Convex Reflector}
\label{sec:fmodel_s3}
We replace the planar reflector with a large convex reflector centered at $x=25$ mm and $z=150$ mm with a radius of 40 mm. A heatmap of the electric field intensity of the FEM simulation for this configuration is shown in Fig~\ref{fig:fem}(d). After impinging upon the circular reflector, the lobes reflect as directional beams, exhibiting both spatial divergence and an increase in beamwidth. 

To capture this behavior in the prediction, we characterize the beam using the same approach as in Section~\ref{sec:fmodel_s1_m}, delineating the left and right edges of the first two lobes. The model takes as input these edges, along with the aperture and $\mathcal{R}(x)$. For each trajectory input, the model produces an output envelope that approximates a linear function. We plot these four linear functions in white over the heatmap in Fig.~\ref{fig:fem}(d). The individual tangents are omitted for clarity. 

The model successfully predicts three key effects induced by the convex reflector: (1) the reflected lobes are directional, (2) the reflected lobes exhibit increased spatial divergence from one another, and (3) the reflected lobes exhibit a diverging beamwidth as they propagate. This scenario highlights the model's versatility in predicting complex reflection behaviors across nonplanar reflector geometries. 

\section{EXPERIMENTAL EVALUATION} \label{experiments}
\subsection{Platform and Curved Beam Generation}
The first step is generating a curving beam at sub-THz frequencies. While beam generation is not the primary contribution of this work, it is necessary to characterize the reflected beam accurately.

\begin{figure}[tb]
    \centering
    \includegraphics[width=\columnwidth, height=2.5in, keepaspectratio]{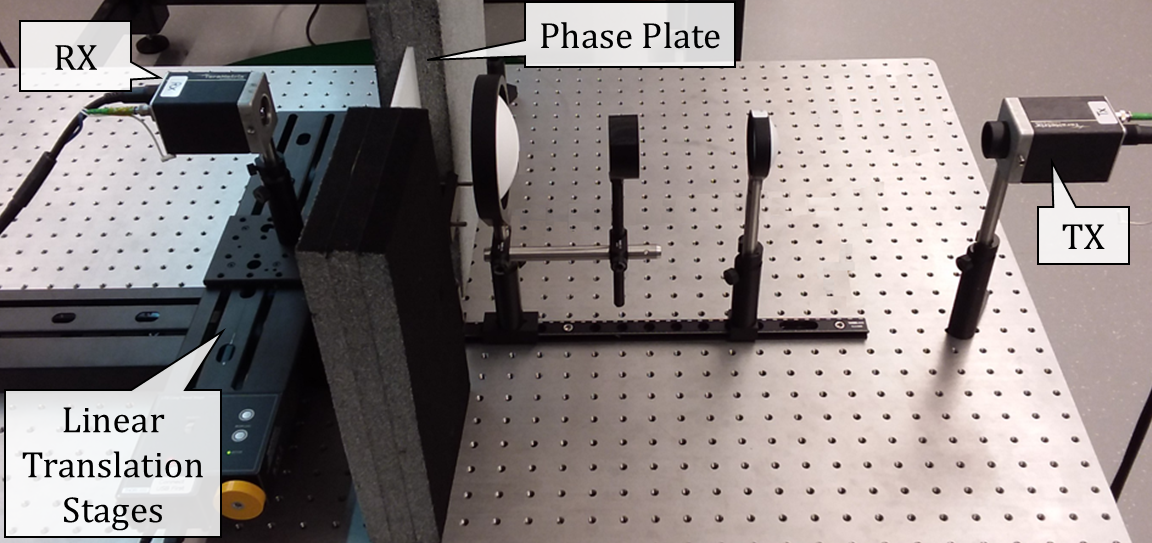}
    \caption{Incident curving beam generation setup.}
    \label{fig:inc_setup}
\end{figure}

We generate beams using phase-only modulation, following the method outlined in~\cite{guerboukha_curving_2024} and~\cite{froehly_arbitrary_2011}.  We calculate the required spatial phase profile of a curving beam with trajectory $C(z)=0.5z^2$. The required thickness $h(x)$ of the phase plate is determined using the equation $h(x)=\phi(x)\lambda / 2\pi(n-1)$ where $n$ is the refractive index of the material. In this work, we use polylactic acid, which has a refractive index of $n=1.6$ at THz frequencies~\cite{guerboukha_curving_2024}. The phase plate is fabricated using an additive 3D printer.

We generate a THz pulse using a broadband LUNA T-Ray 5000 Time-Domain Terahertz Spectroscopy (TDS) system that transmits femtosecond pulses with a bandwidth extending to nearly 3 THz. The TDS receiver mounted on two orthogonally configured motorized linear stages collects the pulses and averages them 100 times. Each linear stage has a 300 mm range. We set the $x$-axis step size to 2 mm and the $z$-axis step size to 5 mm. This allows us to scan and reconstruct the electric field of the curving beam in the $xz$-plane. We show a photograph of the experimental setup in Fig.~\ref{fig:inc_setup}.

\begin{figure} [tb]
    \centering
    \includegraphics[width=\columnwidth]{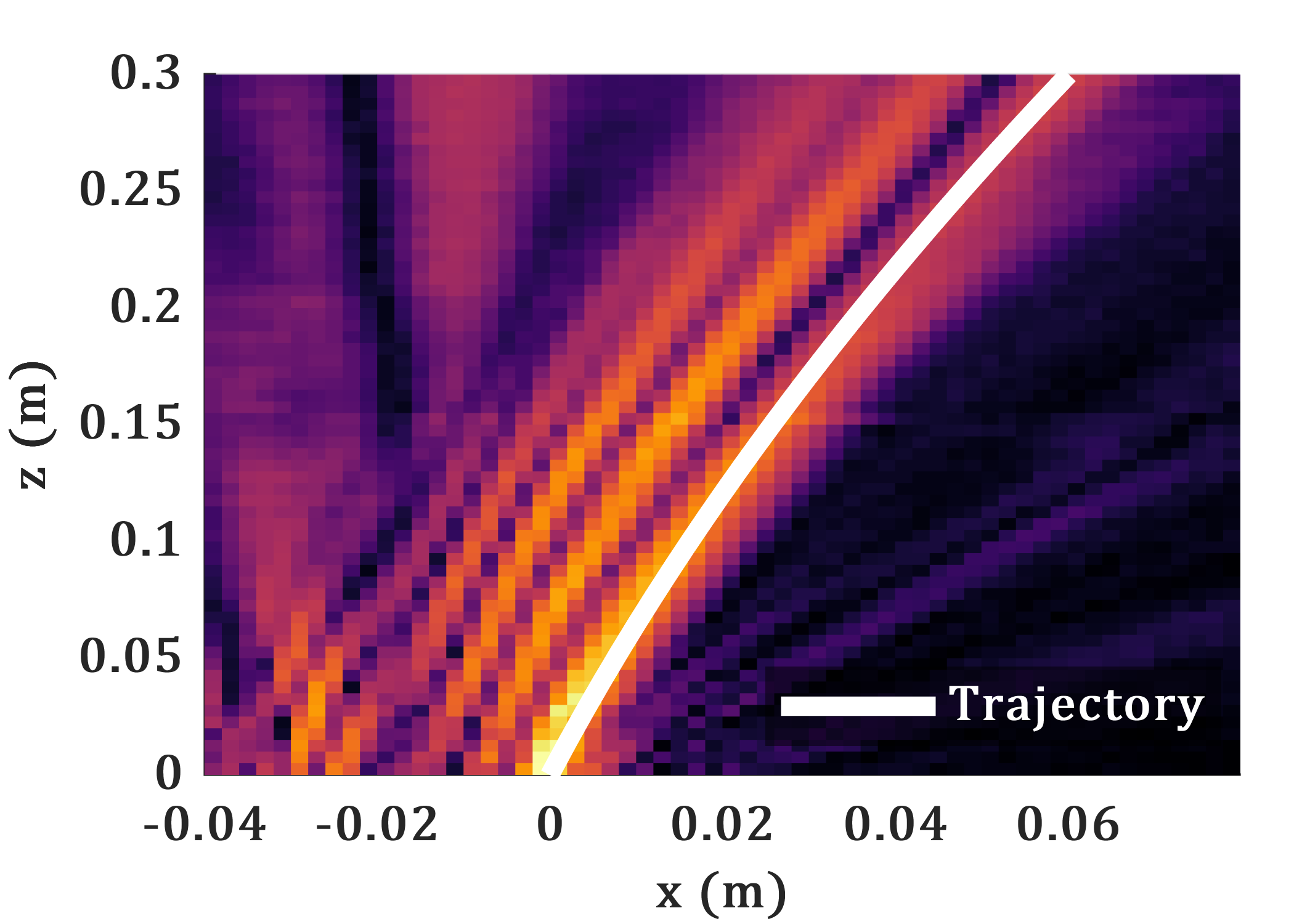}
    \caption{Heatmap of the incident beam at 160 GHz.}
    \label{fig:incbeam_heatmap}
\end{figure}

We present the resulting heatmap of the magnitude of the fast Fourier transform (FFT) at each receiver position at 160 GHz in Fig.~\ref{fig:incbeam_heatmap}. The primary lobe of the beam originates from $x=0$ and exhibits a parabolic trajectory. By fitting a curve to the peak intensity of this lobe, we determine its trajectory to be $\mathcal{E}_i(z)=0.1911z^2+0.0286z-0.0256$ for $z\ge0$, which we overlay in white onto the heatmap for reference. In addition to the primary lobe, secondary lobes emerge following a similar curved trajectory. Notably, the beam's trajectory deviates from the intended design of the phase plate. This discrepancy is largely due to experimental factors, most notably errors from additive 3D printing, where sub-millimeter gaps in the filament can occur during fabrication. Despite this deviation, the beam's overall behavior is curving, which is the primary focus of our analysis. These errors do not affect the model, provided they are known and characterized.

\subsection{Experimentally Measuring the Reflection}
The transmitter, beam expander, and phase plate remain unchanged from the initial setup. We employ an off-the-shelf 12-inch by 24-inch, 22-gauge aluminum sheet for the reflector. The reflector's dimensions are sufficiently large relative to the beam's aperture to produce a well-formed curved reflection. We position the reflector 240~mm away from the phase plate. Prior to the reflection measurements, we conducted preliminary simulations and experimental tests to confirm that the reflector's material, thickness, and roughness were adequate to produce a strong specular reflection at sub-THz frequencies.

\begin{figure} [tb]
    \centering
    \includegraphics[width=\columnwidth]{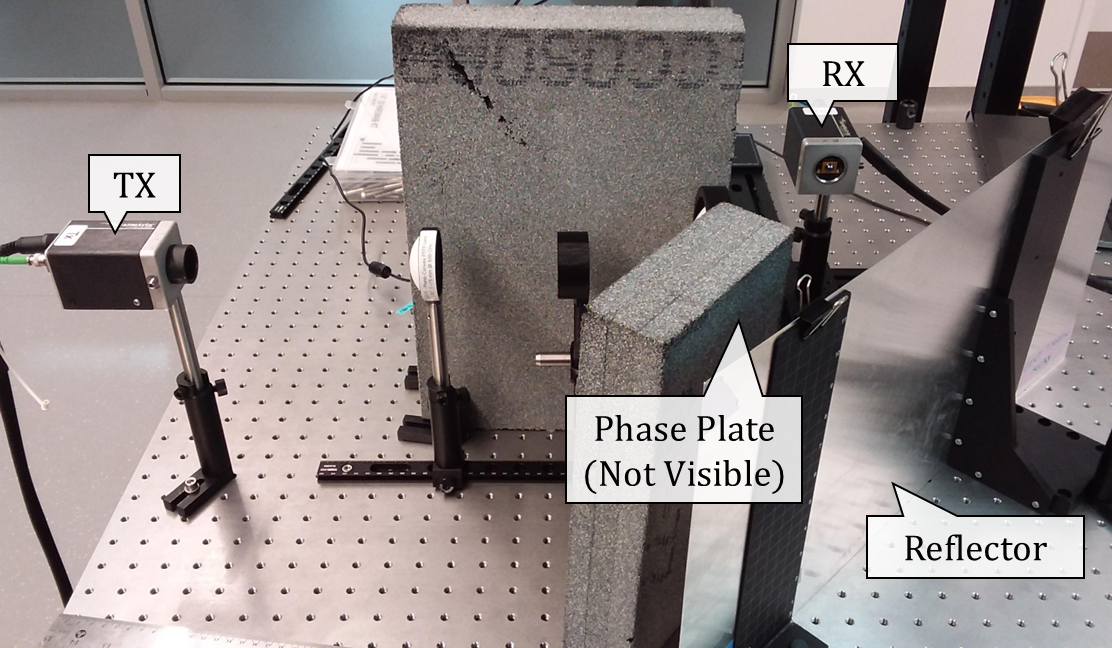}
    \caption{Reflection experiment setup.}
    \label{fig:ref_setup}
\end{figure}

To capture the reflected beam, the TDS receiver and linear translation stages are repositioned along the expected trajectory of the reflection. The receiver can unintentionally block the transmitter if placed too close to the reflector; therefore, the receiver and linear stages are placed 240 mm away from the reflector. The receiver is oriented orthogonally to the transmitter to achieve the best average power across the entire reflection. The $x$-axis is transverse to the phase plate aperture, and the $z$-axis is the propagation axis of the incident beam. We decrease the step size from the previous experiment to 2 mm in the $x$-axis and 2 mm in the $z$-axis. We provide a photograph of this experimental setup in Figure~\ref{fig:ref_setup}.

\begin{figure} [tb]
    \centering
    \includegraphics[width=\columnwidth]{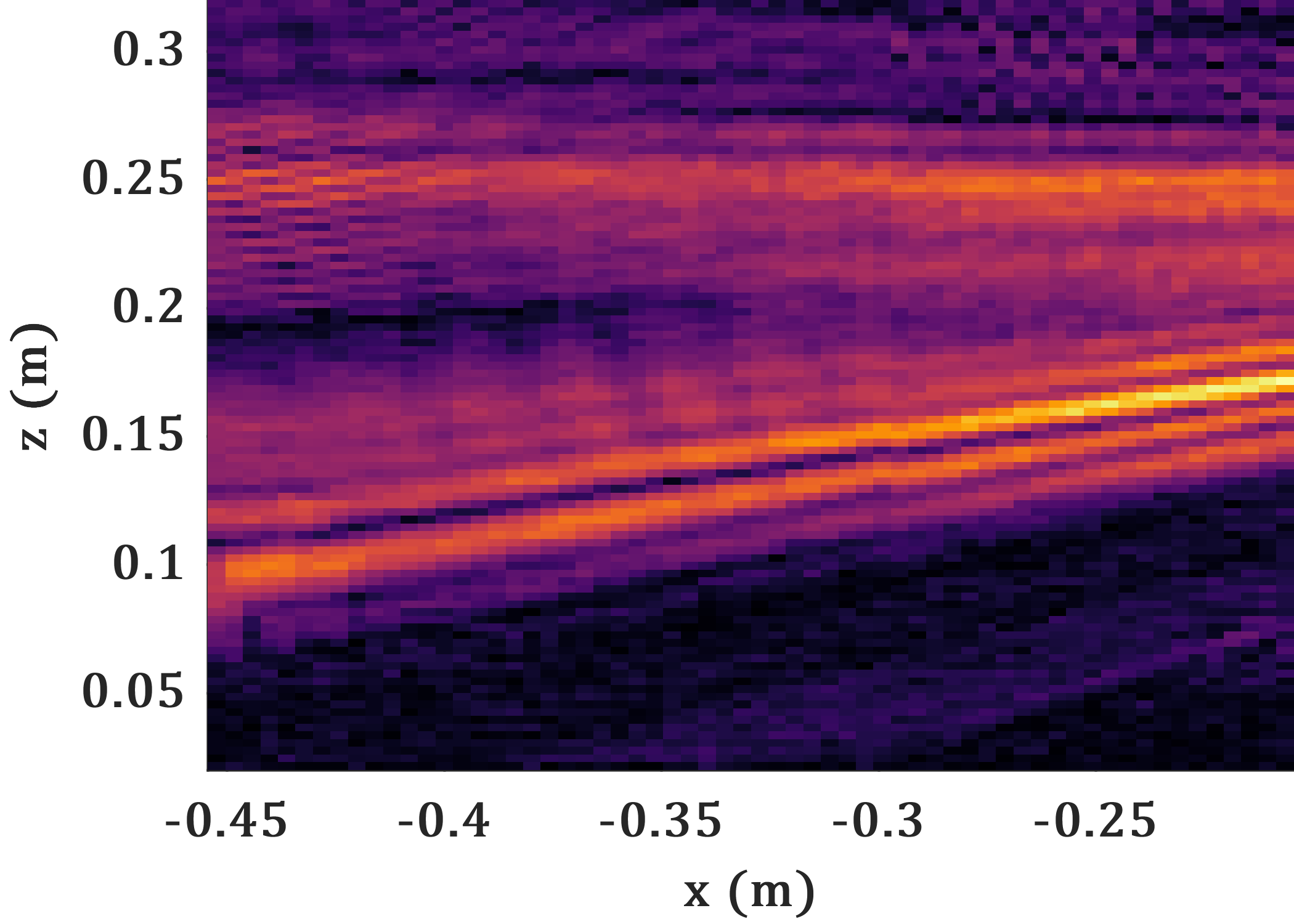}
    \caption{Experimental curved beam reflection.}
    \label{fig:exp_raw}
\end{figure}

\textbf{Results.}
We compute the magnitude of the FFT at 160 GHz and present the resulting heatmap of the reflected beam in Figure \ref{fig:exp_raw}. We find the equation of best fit for the peak of the first lobe to be $\mathcal{E}_r(x)=-0.0161x^2 + 0.3068x + 0.2104$. Note that the curvature of the reflection appears modest due to the measurement distance (over 240 mm away from the reflector) chosen to avoid obstructing the incident beam. The lobes maintain their width post-reflection, though the first lobe diminishes in intensity more rapidly than the second and third lobes. This is a result of both the phase-only beam generation and the orientation of the receiver.

Beyond the third lobe, the reflection is less structured, with no distinguishable lobes. This phenomenon is consistent with our prior simulations, which show that a curving beam with an excessive truncation factor ($a>0.1$) incident upon a large planar reflector will result post-reflection in a similar loss of structure, as seen in Fig.~\ref{fig:fem}(a). However, the persistence of several distinct curving lobes post-reflection aligns with our simulation results for a large planar reflector. These findings confirm that the reflected beam's trajectory and overall behavior are consistent with both FEM simulations and the general predictions of our model. 

\subsection{Predicting the Experimental Reflection}
In this section, we evaluate our model's ability to predict the reflection of the experimental curving beam. We input the equation describing the trajectory of the incident beam~$\mathcal{E}_i(z)=0.1911z^2+0.1430z$, the equation of the reflector~$\mathcal{R}(x)=-x+0.238$, and the aperture size $x_a=0.1$, which corresponds to the physical dimensions of the phase plate. Using this information, the model predicts that the reflected trajectory of the main lobe along its peak will be $\hat{\mathcal{E}}_r(x)=-0.1090x^2 + 0.2442x + 0.1930$. 

\begin{figure}[tb]
    \centering
    \includegraphics[width=\columnwidth, height=2.5in]{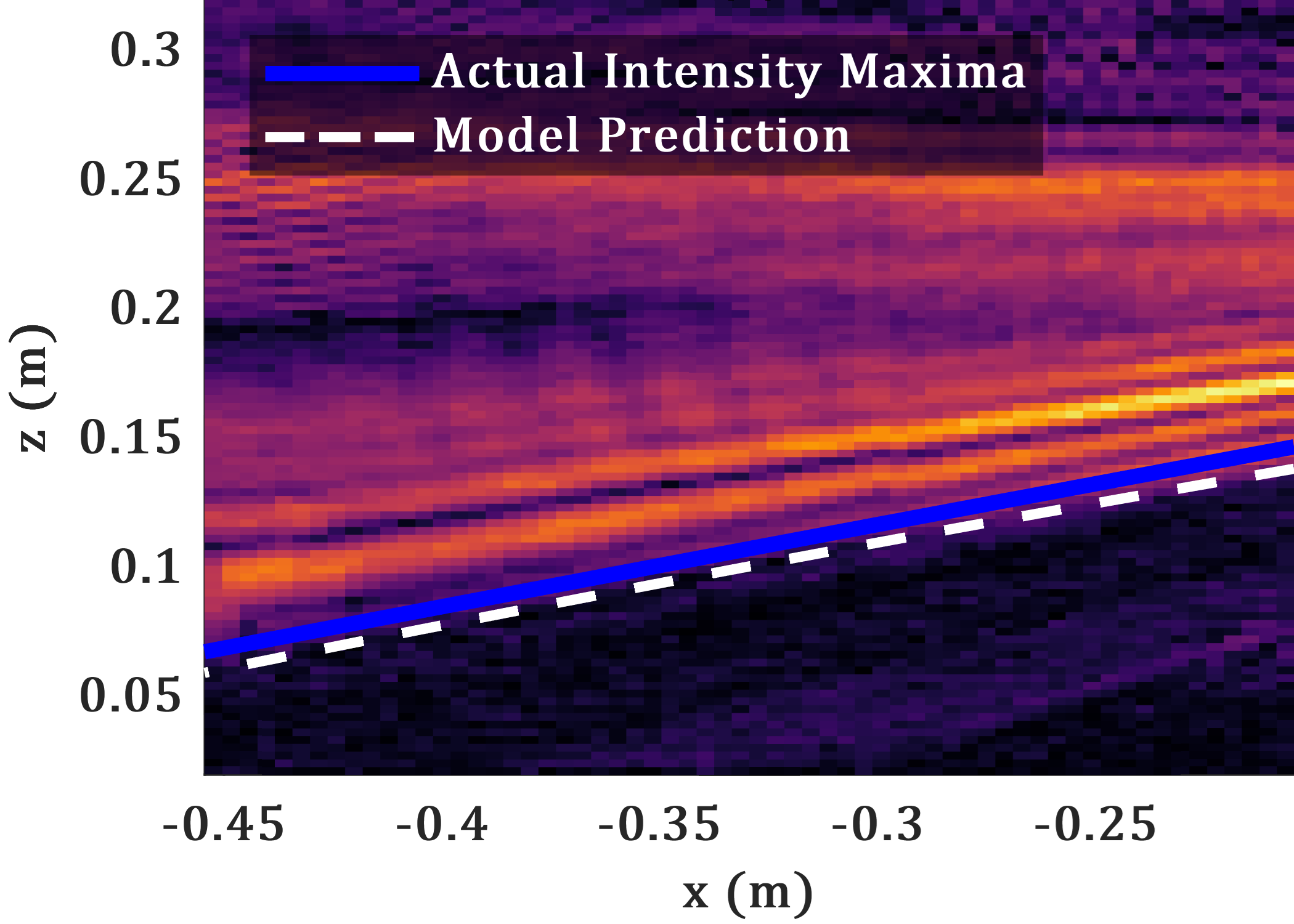}
    \caption{Predicting the intensity trajectory of the first lobe.}
    \label{fig:exp_ref}
\end{figure}

To assess the accuracy of this prediction, we overlay the model's output onto the experimentally measured electric field data, as shown in Fig.~\ref{fig:exp_ref}. The model successfully predicts that the reflection off the large planar reflector will curve, but at 240 mm away from the reflector, this curvature will appear modest. We find the RMSE to be 7.4~mm, which is less than four wavelengths. Additionally, the maximum distance between the predicted and experimental trajectories is 8.5~mm. These results indicate that while the model provides a strong prediction of the reflected beam's trajectory, some discrepancies do exist. Specifically, the prediction seems to be accurate regarding the curvature of the reflected trajectory but has a translational error in the $z$-axis.

Comparing this performance to FEM simulations (Section~\ref{sec:fmodel_s1}), where the RMSE was reported to be 1.4~mm, we observe a higher degree of error in the experimental setting. Several factors likely contribute to this increased deviation, including imperfections in the reflector’s surface and our assumption that the beam is invariant in the $y$-direction, where, experimentally, this is not the case. Future extensions to the model may benefit from incorporating analysis in the third dimension, which may further increase accuracy. Despite these sources of error, the model successfully captures the behavior of the reflection using a single function.

\begin{figure}[tb]
    \centering
    \includegraphics[width=\columnwidth]{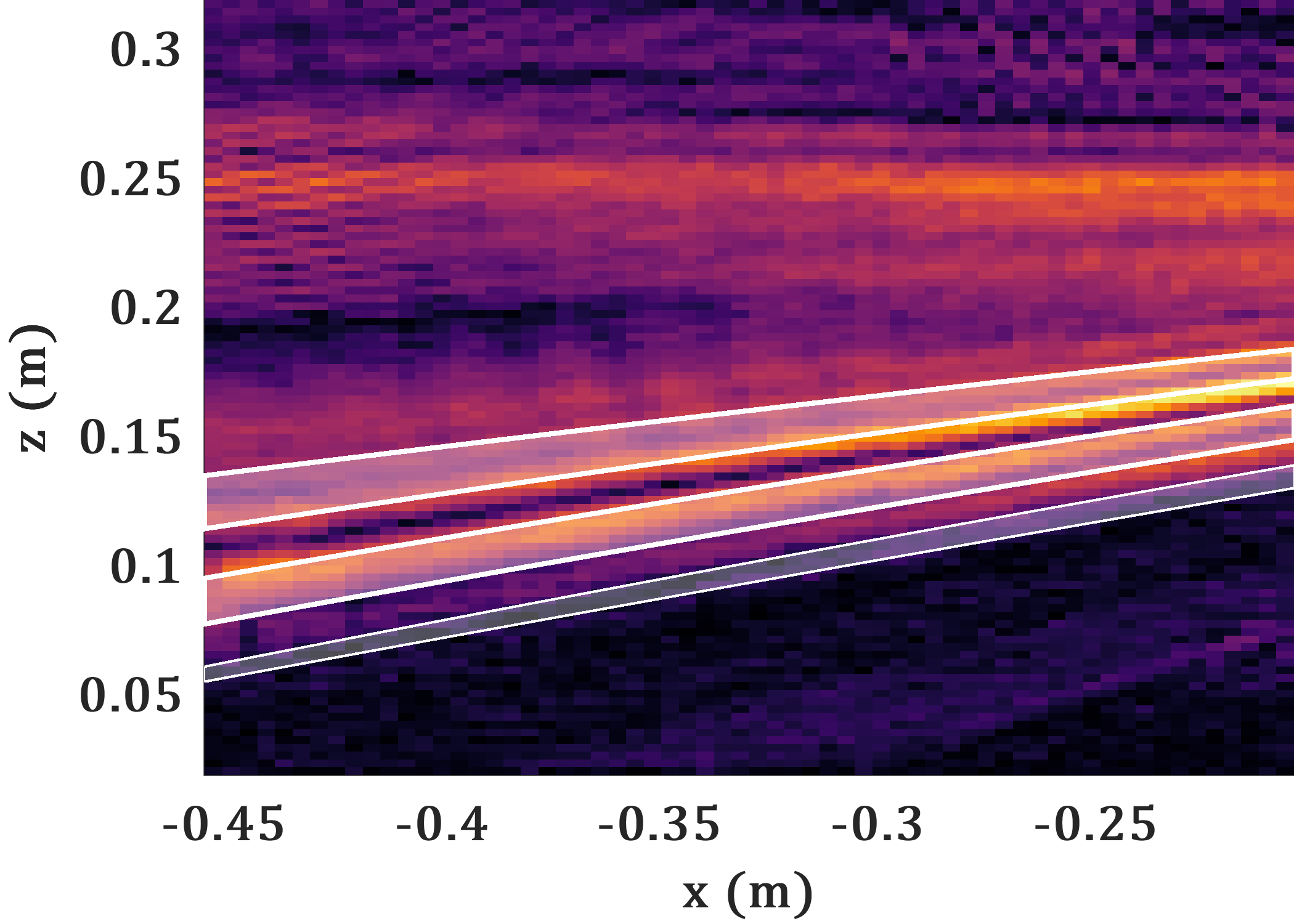}
    \caption{Predicting the trajectory of the first three lobes of the reflection.}
    \label{fig:exp_lobes}
\end{figure}

We also perform a prediction of the widths of the first three lobes of the incident beam, similar to our analysis in Section~\ref{sec:fmodel_s1_m}. We display the results in Fig.~\ref{fig:exp_lobes}. We find that the model successfully predicts that the three lobes will each follow a curving trajectory post-reflection; however, there again exists a slight translational error of several millimeters consistent with the single-trajectory prediction.

\section{RELATED WORK}
\label{relatedwork}
\subsection{Curving Beams in Free Space} 
In 1979, Berry and Balazs demonstrated that the time-dependent Schrödinger equation admits a solution in terms of the Airy function, implying the existence of a wave packet that self-accelerates~\cite{berry_nonspreading_1979}. An electromagnetic wave analogue of this phenomenon was experimentally demonstrated in 2007, named the Airy beam~\cite{siviloglou_observation_2007}. Subsequent research revealed that Airy beams are a special case within a broader class of self-accelerating beams, which can take on a variety of shapes~\cite{froehly_arbitrary_2011, greenfield_accelerating_2011}. Most prior work on these beams has remained within the optical domain, focusing on their generation, characterization, and applications such as particle manipulation and micromachining, among others~\cite{zhang_advances_2016, baumgartl_optically_2008, mathis_micromachining_2012}. Only recently have these beams been proposed for wireless systems at THz frequencies~\cite{singh_wavefront_2024, guerboukha_curving_2024, lee_experimental_2025}, raising new questions that remain unresolved even in optics, one of which concerns reflection. 

\subsection{Curving Beam Electric Field at an Infinite Dielectric Plane} 
Previous research on the reflection of curving beams has primarily focused on analyzing the electric field characteristics of Airy beams at dielectric interfaces~\cite{yang_reflection_2022, chremmos_reflection_2012, yang_characteristics_2022}. One study introduced a vector wave framework to examine the reflection and refraction of 2D Airy beams~\cite{hui_vector_2020}. However, in wireless communication, sensing, and imaging applications, the interest extends beyond Airy beams to a broader class of curving beams, necessitating an understanding of reflections from finite and arbitrarily shaped reflectors. Our work takes a distinct geometric approach that generalizes beyond Airy beams to convex beam trajectories and enables the prediction of reflections from nonplanar and finite reflectors - factors that significantly impact beam reflection.

\section{CONCLUSION}
\label{conclusion}
In this work, we propose and experimentally validate a geometric model for predicting the reflection of convex beams off arbitrary reflectors. We demonstrate that conventional ray tracing approaches fundamentally cannot account for curving beam propagation and that attempting to ``mirror'' the trajectory of the normal of the reflector fails in general. We design our model around the Legendre transform, which allows us to represent the trajectory as a family of tangents. By applying the laws of reflection to these tangents individually and reconstructing the reflected trajectory, we successfully captured the behavior of curving beams upon reflection, even from finite or convex reflectors. 

We evaluate the model with a combination of FEM simulation and experimental measurements, showing that we can predict the reflection of a curving beam with errors as low as 1.4 mm and 7.4 mm, respectively. 
This work bridges a key gap in sub-THz wireless design. Though often overlooked in physics, curving beam reflections are critical for emerging communication and sensing systems, where channel behavior will be strongly affected by reflector geometry. Future work includes studying and incorporating electromagnetically rough reflectors into the model.

\section*{Acknowledgments}
An earlier version of this work appeared in~\cite{spindel_caroline_j_curving_2025}.
This research was supported by Cisco and Intel, by NSF grants 2433923, 2402783,
and 2211618, and by ARO DURIP grant W911NF-23-1-0340.

\bibliographystyle{IEEEtran}
\bibliography{refs}

\vspace{12pt}

\end{document}